\documentclass[runningheads]{apinv}
\usepackage{epsfig,graphics, natbib}
\usepackage{marvosym} 

\usepackage[utf8]{inputenc}

\begin{document}

\title{ The origins of the magnetic field in tip-RGB, AGB and post-AGB stars}
\titlerunning{Magnetic field in tip-RGB, AGB and post-AGB stars}
\author{Renada Konstantinova-Antova\inst{1}, Stefan Georgiev\inst{1,2}, Agn\`es L\`ebre\inst{2}, Corinne Charbonnel\inst{3,4}, Ana Palacios\inst{2}, Michel Auri\`ere\inst{4} }
\authorrunning{R.Konstantinova-Antova et al.}
\tocauthor{Renada Konstantinova-Antova, Stefan Georgiev, Agnes Lebre et al.} 
\institute{Institute of Astronomy and NAO, Bulgarian Academy of Sciences, BG-1784, 72 Tsarigradsko shosse, Sofia, Bulgaria
	\and LUPM, University of Montpellier, CNRS, Montpellier, France
	\and Department of Astronomy, University of Geneva, Chemin des Maillettes 51, CH-1290 Versoix, Switzerland
	\and IRAP, Universit\'e de Toulouse, CNRS, Observatoire Midi Pyr\'en\'es,  57 
Avenue d'Azereix, 65008 Tarbes, France   \newline
	\email{renada@astro.bas.bg}    }
\papertype{Review, Submitted on xx.xx.xxxx; Accepted on xx.xx.xxxx}	
\maketitle

\begin{abstract}
During the last decade and a half, the new generation spectropolarimeter Narval at Pic du Midi, France allowed the study of weak magnetic fields in cool giant stars that are fairly evolved after main sequence. We present a short summary on the recent knowledge on the magnetic fields and activity in giants situated in the upper right part of the Hertzsprung-Russel (H--R) diagram and discuss on the possible mechanisms for magnetic field generation in the asymptotic giants branch (AGB) and post-AGB stars.
\end{abstract}

\section*{Introduction}
When a star evolves after the main sequence, it  expands. A convective envelope begins to develop in intermediate mass stars, and in low-mass stars the existing one deepens toward the stellar core. The stars ascending the red giant branch (RBG) possess vigorous convection in a deep convective envelope. Their internal structure is a complex one, consisting of a hydrogen burning shell and a contracting core. After the star reaches the tip of the RGB, the He-burning follows. Later on, after the He--burning phase, the star continues to climb up and right on the Herzsprung-Russel (H--R) diagram, on the asymptotic giant branch (AGB) and its structure becomes even more complex, with hydrogen and helium burning shells and contracting core.

During the course of evolution, at certain phases additional angular momentum as a result of core-envelope interaction or eventual planet engulfment could speed up the stellar rotation and operation of an $\alpha$ - $\omega$ dynamo could become possible there \citep{schroderrenada2022}. Also, when the star crosses the instability strip, pulsations begin to occur in its envelope and propagate trough the atmosphere. They could compress the existing weak magnetic field (MF) at certain phases of the pulsation.

We present here the first findings on how all these structural changes in tip-RGB, AGB and post-AGB stars are related with their magnetic fields and activity. This knowledge is based on more than 15 years of study of the magnetic fields and activity in such stars in the framework of a Bulgarian-French collaboration using the spectropolarimeter Narval at TBL, Pic du Midi, France \citep{auriere2003} and the Least Square Deconvolution (LSD) method \citep{donati97}, that enabled detection of weak magnetic fields of the order of 1~G and below in cool stars, as a result of averaging more than 10~000 absorption lines in their spectrum.



\section{First studies of magnetic field in tip-RGB and AGB stars. Discovery of the second magnetic strip on the H-R diagram where $\alpha - \omega$ dynamo operates. }
\label{sec:first-studies}
Until a decade ago, magnetic fields in evolved giants  situated at the tip of the RGB, AGB and beyond on the H--R diagram were poorly studied. Some suspicions that M giants could be magnetically active appeared in \cite{hunsch1998} on the basis of enhanced X-ray emission in four M giants. Herpin et al. (2009) found evidence for a MF in the circumstellar envelopes of four AGB stars on the basis of IRAM-30m observations. From a theoretical point of view, \cite{soker2000}, \cite{sokerzoabi2002}, \cite{brandenburg2002} and \cite{nordhaus2008} predicted that magnetic dynamo of different types could operate in AGB stars.

\begin{figure}
    \centering
    \includegraphics[width=\columnwidth]{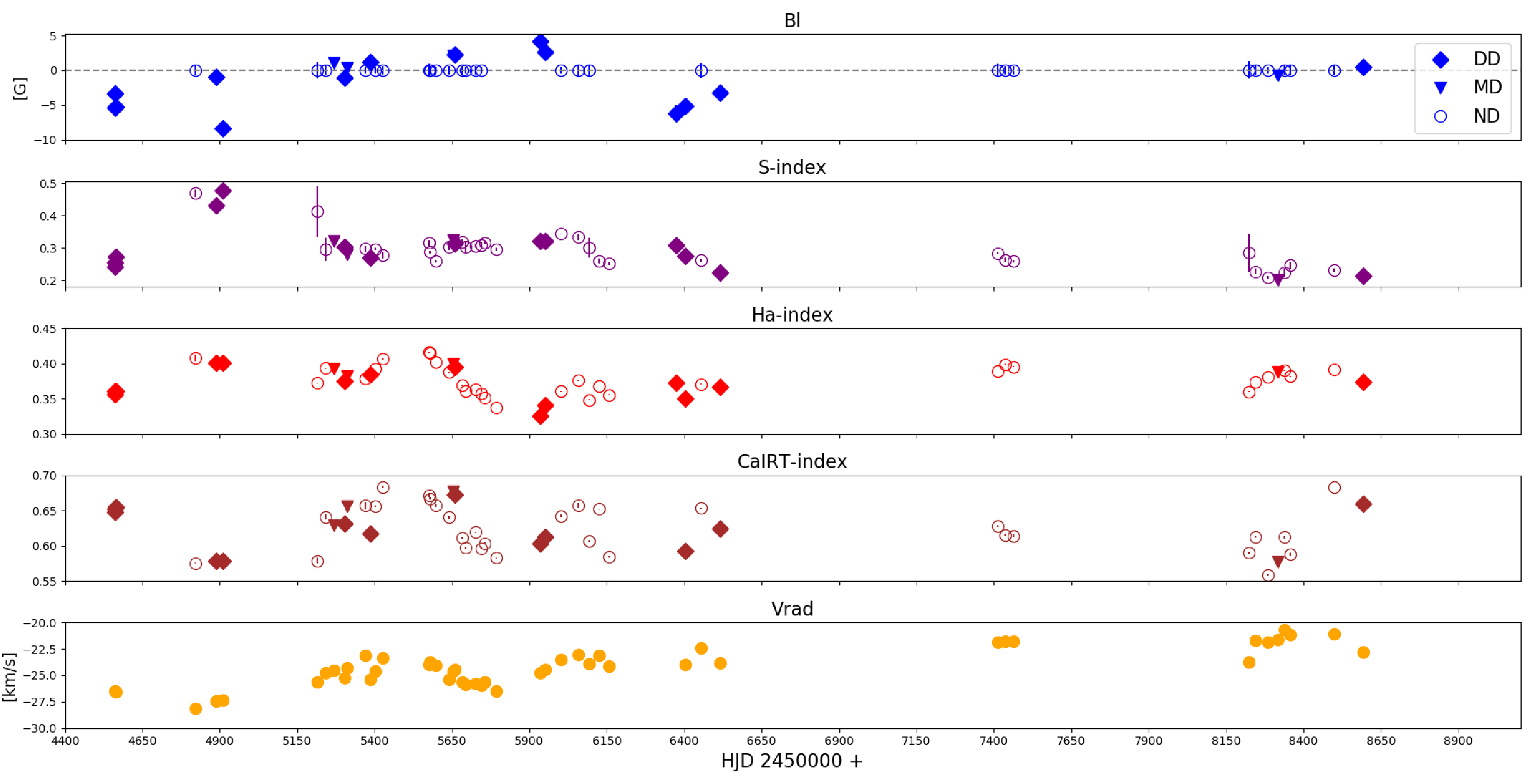}
    \caption{Longitudinal magnetic field $B_{l}$ and line activity indicators variability in EK~Boo. Periods of MF  detection and non-detection are visible (filled and open symbols, respectively). Radial velocity long-term variation indicates for a companion at a distant orbit. Such one is not visible in the stellar spectrum and is likely a  dwarf. Figure from \cite{georgiev2020}.}
    \label{fig:ekboo-var}
\end{figure}

We started our study with a sample of 9 apparently single M giants selected for their faster rotation and/or X-ray emission in 2008. The first Zeeman-detected M giant is EK~Boo = HD~130144, a fast rotator ($V\sin i$ = 8.5 \kms) with an unusual X-ray emission. A longitudinal magnetic field ($B_l$, the magnetic field component projected along the line of sight) reaching up to 8~G was measured for it \citep{renada2010}. Later on, 6 more M giants from this sample were also Zeeman-detected and weak magnetic fields of the order of 1~G were measured for them with one exception, the M5 giant RZ~Ari, which possesses a magnetic field even stronger than EK~Boo \citep{renada2013}. All these objects are tip-RGB or AGB stars \citep{renada2010}.  An interesting fact for them is that their magnetic field was not easily detected, because they display both periods of detection and periods of non-detection, contrary to the magnetic G and K giants, that begin to climb the red giant branch (RGB) or those in the He-burning phase \citep{auriere2015}. A typical example of a star exhibiting alternating periods of detection and non-detection is EK~Boo. Its long-term MF and activity indicators behavior is presented in Figure~\ref{fig:ekboo-var} \citep{georgiev2020}. In this figure, different symbols represent the different levels of magnetic field detection based on a reduced $\chi^2$ test calculated from the LSD profiles as described in \citep{donati97}. The results of this test are converted into a False Alarm Probability (FAP), based on which three different levels of detection are introduced: definite detection (DD, FAP~$<~10^{-4}$), marginal detection (MD, $10^{-2} >$~FAP~$>~10^{-4}$) and no detection (ND, FAP~$>~10^{-2}$. These periods of MF detection and non-detection in the studied M~giants could be a result of the convective cells lifetime \citep{freytag17}. These cells are believed to be the transporter of the magnetic field.

\begin{figure}
    \centering
    \includegraphics[width=\columnwidth]{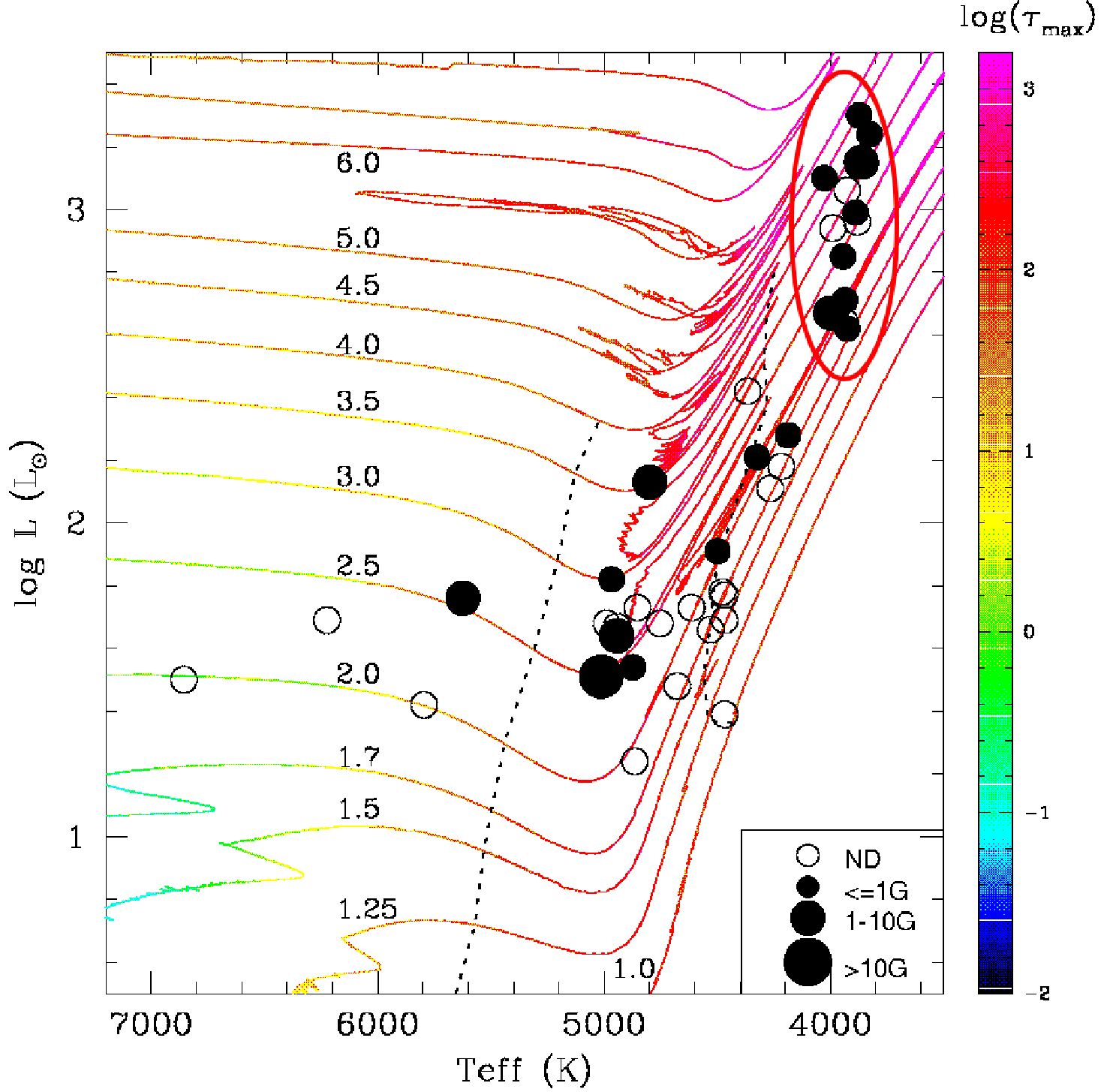}
    \caption{Situation of the Zeeman detected giants from Solar vicinity sample on the H-R diagram. Longitudinal magnetic field $B_{l}$ strength is designated by symbols with different size. Open circles designate stars with no detection. Dashed lines mark the first dredge-up phase. Colors stand for the maximal convective turnover time at each point of the evolutionary tracks. Figure from \cite{renada2014}. The locus of the second "magnetic strip" is designated by an ellipse. }
    \label{fig:magnetic-strip-2}
\end{figure}

However, this first sample is a biased one and cannot answer the question of what percentage of the tip-RGB and AGB stars are magnetically active. For this purpose we decided to study all such stars up to 4th magnitude in V-band in the Solar vicinity which are not binaries and are available for observations at Pic du Midi observatory, France. We limited the brightness up to 4th magnitude because we wanted to be able to perform a deep study, down to 0.2~G accuracy. In this way, we studied 17 tip-RGB and AGB stars and about 60 percent of them were Zeeman-detected \citep{renada2014}. Weak magnetic fields of a few Gauss and even at the sub-gauss level were detected for them. It appeared that all these stars form a strip on the H--R diagram which we called “second magnetic strip” (Figure~\ref{fig:magnetic-strip-2}). The first magnetic strip is observed in G and K giants, situated at the base of the RGB and in the He-burning phase \citep{auriere2015}.  Later on, \cite{charbonnel2017} show with evolutionary models that a dynamo of the $\alpha - \omega$ type could be the explanation for the M giants clumping in a strip (and earlier, in the first magnetic strip for the G and K giants situated at the base of the RGB and the He-burning phase). They found that these areas on the H--R diagram are favorable for the $\alpha - \omega$ dynamo, because of the conditions in the convective envelopes, in combination with the stellar rotation that enable a Rossby number, $R_o$ ($R_o$ =  $P_{rot}$/$\tau_{conv}$, where $P_{rot}$ is the rotation period of the star and $\tau_{conv}$  is the convective turnover time) less than one there and, hence, an efficient action for this type of dynamo.

\section{Cool giants with magnetic activity outside the magnetic strips }  

Besides the case of $\alpha - \omega$ dynamo operation in giants situated in the first and second magnetic strips and depending mostly on rotation and the convective envelope properties (like the convective turnover time, $\tau_{conv}$), a few fairly evolved giants outside the magnetic strips were found. These are mostly AGB and post-AGB stars with a magnetic field of a few Gauss. Their magnetic field cannot be explained by the  $\alpha - \omega$ dynamo, because of the big Rossby number, $R_o$. A different type of dynamo, or even different mechanism for the magnetic field production, could operate in these giants.

\subsection{Pulsating magnetic giants }

Some AGB and post-AGB stars are known to exhibit radial pulsations, which generate strong shockwaves - mechanical waves with velocities larger than the sound velocity in the given medium. These shockwaves propagate through the stellar atmosphere, causing a sharp change in temperature, pressure and density at the shock front. As the shock propagates, it creates complex ballistic motions in its host environment, causing different layers to move with different velocities and in different directions with respect to the stellar restframe, which in turn affects the spectral lines that form there. Layers affected by the shock are initially caused to move away from the photosphere and towards the observer, resulting in blueshifted (with respect to the restframe of the star) spectral lines; after sufficient amount of time has passed since the shock's appearance, these same layers then begin to fall ballistically towards the photosphere and away from the observer, resulting in redshifted spectral lines.

\begin{figure}
    \centering
    \includegraphics[width=\columnwidth]{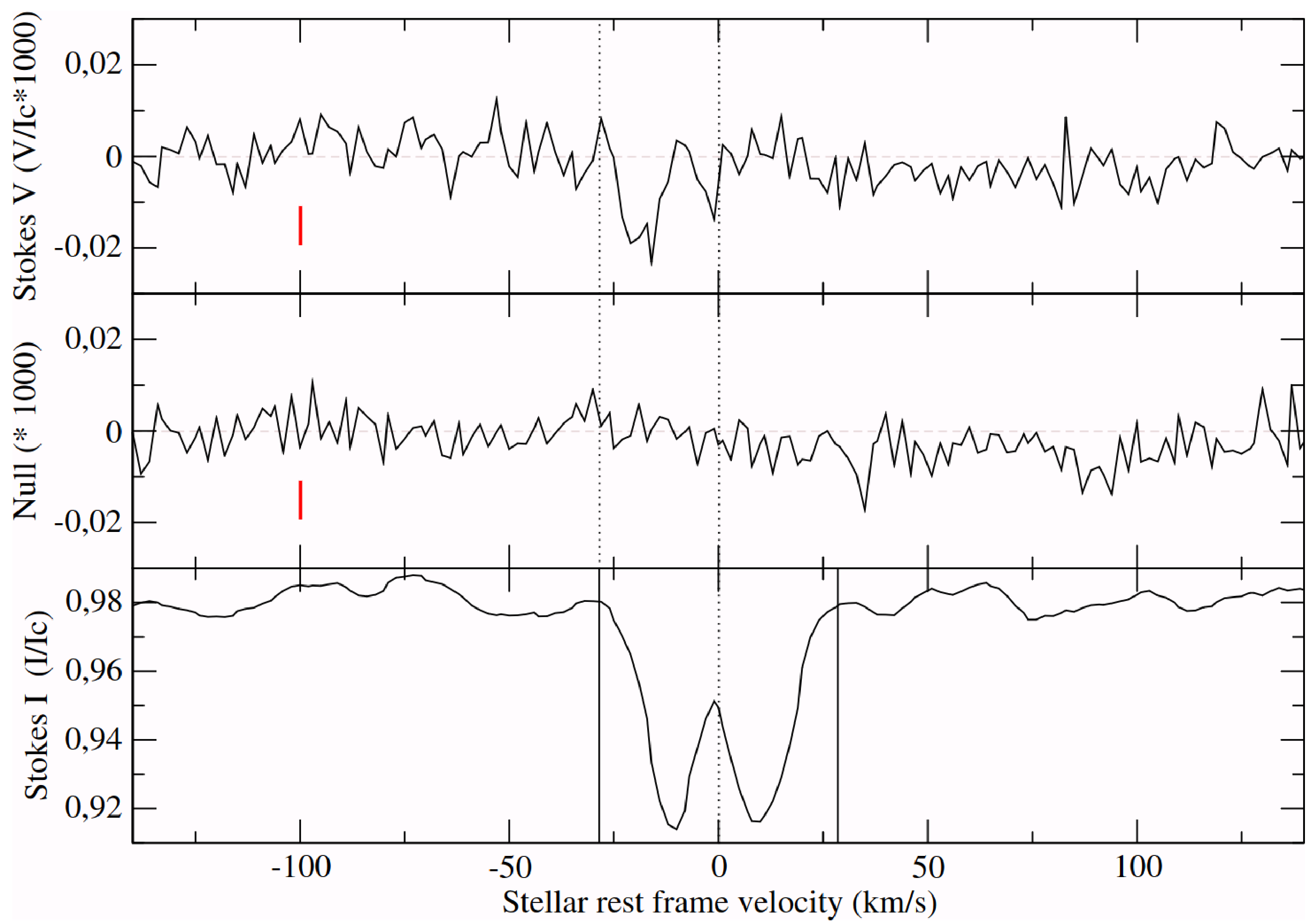}
    \caption{LSD profile of $\chi$~Cyg, computed from Narval observations taken in March 2012. From bottom to top, the three panels in the figure show the intensity profile (Stokes~$I$), the diagnostic null profile and the circularly polarized profile (Stokes~$V$). It can be seen that the signal in the Stokes~$V$ profile is clearly associated to the blue wing of the intensity profile. Figure from \cite{lebre2014}.}
    \label{fig:chicyg-lsd}
\end{figure}

The first detection of a surface magnetic field as a result of shockwave propagation was in a Mira star (in the AGB evolutionary stage) was reported by \cite{lebre2014}. They measured a longitudinal magnetic field $B_l$ of 2-3~G for the star $\chi$~Cyg from spectropolarimetric observations with Narval. \cite{lebre2014} found that in $\chi$~Cyg, the circularly polarized signatures found in the LSD profiles obtained from high resolution spectra of the star are clearly associated to the blue lobe of the atomic spectral lines, i.e. are being accelerated by the shockwave. From this observation, the authors concluded that the shockwaves propagating in the atmospheres of Mira stars may be locally amplifying the weak surface magnetic field already present in the affected atmospheric layers. An example is shown in Figure~\ref{fig:chicyg-lsd} (from \citealt{lebre2014}), where the LSD profile of an observation of $\chi$~Cyg near maximum light is displayed, and it can be seen that the Stokes~V profile is clearly associated to the blueshifted lobe, which is related to the matter affected by the shock.

Later, \cite{sabin2015} reported the first detections of a weak magnetic field in two pulsating post-AGB stars, R~Sct and U~Mon. \cite{lebre2015} found indications of the same connection between the observed surface magnetic field and the atmospheric dynamics in R~Sct, which is a pulsating variable of the RV~Tau type. The authors observed R~Sct in circular polarization with Narval and discovered that when circularly polarized profiles were present in the LSD spectrum of the star, they were always associated to the blue lobe of the intensity profile, as in $\chi$~Cyg.


To investigate the hypothesis of a connection between the atmospheric dynamics and surface magnetism in pulsating post-AGB stars, \cite{georgiev2023} present the longest monitoring of the pulsating variable star R~Sct done using high resolution spectropolarimetry. The authors analyze observations of R~Sct obtained with Narval on 67 different nights in the period between July 2014 and August 2019, of which 31 observations contain Stokes~V measurements that allow studying the surface magnetic field. The authors confirm the stellar and Zeeman origin of the observed circularly polarized profiles, and show that their timescale of variation is similar to that of the atmospheric dynamics. Using a new approach for the calculation of LSD profiles for this star, \cite{georgiev2023} constrain the LSD analysis to the lowest parts of the stellar atmosphere and are able to study the longitudinal magnetic field at the level of the photosphere, where the shock is just emerging. The authors found that for all observations where the LSD Stokes~I profile displays a double peak (indicating that the atmosphere is strongly affected by a shockwave), the Stokes~V signature, if present, is associated to the blue intensity lobe in radial velocity space.

\begin{figure} \centering
	\includegraphics[width = \linewidth]{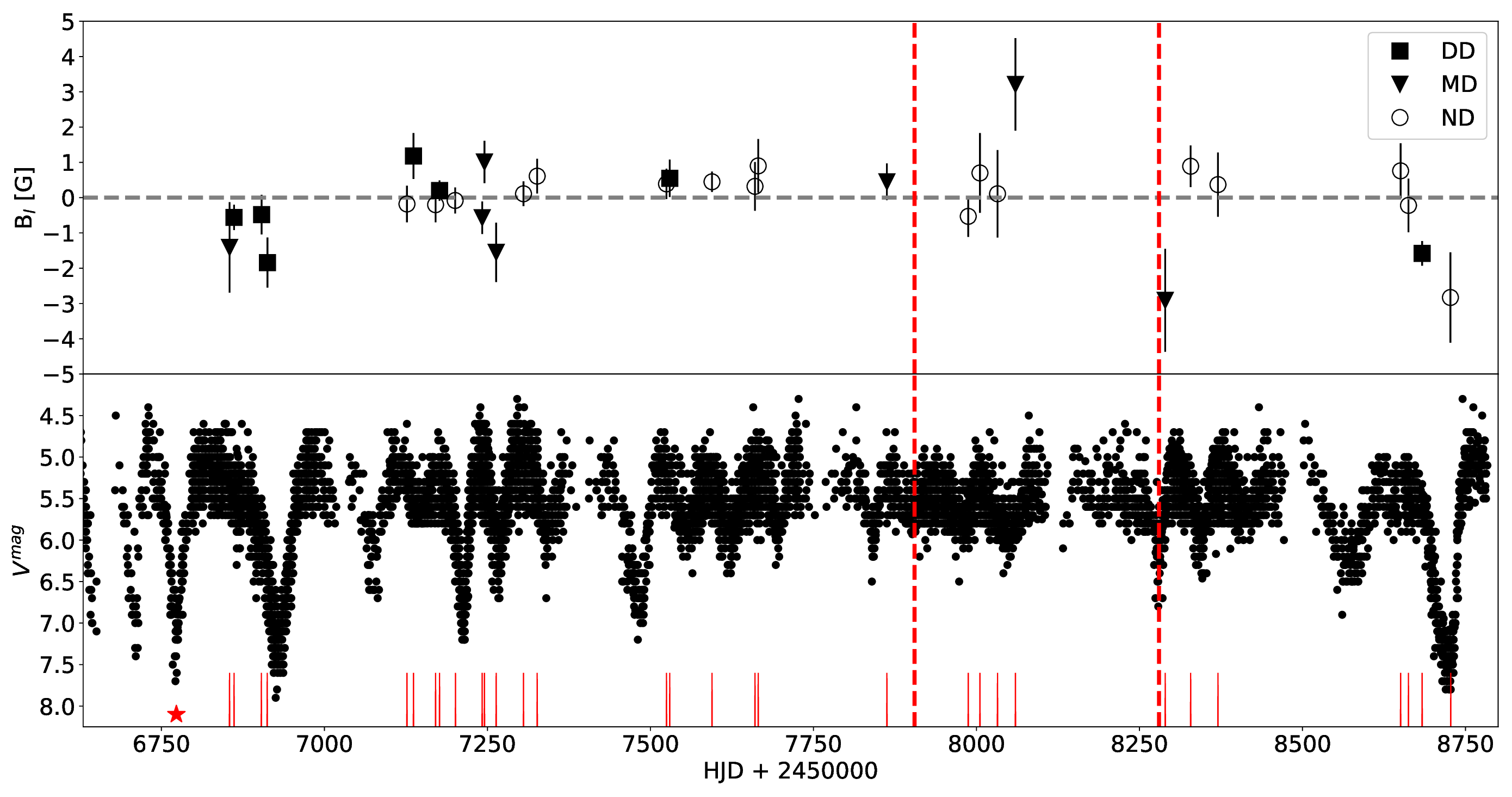}
	\caption[]{Temporal evolution of the longitudinal component of the magnetic field $B_l(t)$  (upper panel) and lightcurve (bottom panel) of R~Sct. In the upper panel, the different symbols represent the different types of detections, described in Section~\ref{sec:first-studies}. The red vertical lines in the bottom panel indicate the time of Narval observations that contain Stokes V measurements, and the red star indicates the date considered as the start of reference period $P_0$ (see Section~2.1 of \citealt{georgiev2023}). The dashed vertical red lines indicate the approximate beginning and end of the irregular part of the lightcurve. Figure and caption adapted from \cite{georgiev2023}}
	\label{fig:rsct_bl}
\end{figure}

In Figure~\ref{fig:rsct_bl} (from \citealt{georgiev2023}), the temporal evolution of $B_l$ of R~Sct between 2014 and 2019 is presented together with the visual lightcurve (from AAVSO). In this figure, it can be seen that the surface magnetic field varies with time. However, there is no clear correlation between the field strength and the photometric phase of the star, suggesting that the variability observed in the longitudinal magnetic field is not entirely caused by the propagation of shockwaves and the resulting local amplification of the surface magnetic field, but also by a variable and non-homogeneous magnetic structure at the surface level of R~Sct.

\subsection{The role of planet engulfment for magnetic field generation }
When a star expands during the giant stages, its radius could reach the orbit of an eventual planet and cause the planet to be engulfed. As a result, an increase of the rotation rate of the star and change in its chemical composition (especially Lithium) occurs. The first idea of the result of planet engulfment in cool giants is presented in \cite{siess1999}. Later, more studies were presented by different authors, among them the detailed models by \cite{privitera2016phd} and \cite{privitera2016a, privitera2016b} who modeled the rotation rate from the base of the RGB up to the AGB for stars of different mass in the case of  planet engulfment. According to them, after this event the giants should rotate faster for some interval of time. The degree to which the rotation speeds up depends on the mass and orbital momentum of the engulfed companion, a planet or brown dwarf, and also on the mass of the star. Since for each star the planetary orbits are situated at different distances, the engulfment could occur at any different evolutionary stage on the giant branches. Hence, the type of dynamo and its efficiency depend on the properties of the convective envelope and the rotation there. In the case where the Rossby number is less than 1, operation of the $\alpha-\omega$ dynamo is possible. Otherwise, other types of dynamo could operate there, like the $\alpha^2-\omega$ one, or the turbulent dynamo (which does not depend on the rotation). The strength of the magnetic field depends on the dynamo type. For example, it was shown by \cite{sokerzoabi2002} that the turbulent dynamo could produce a surface magnetic field of the order of one Gauss in cool giants. In any case, our experience up to now shows that the most effective dynamo in evolved cool giants is the $\alpha - \omega$ dynamo \citep{schroderrenada2022}. The loci on the H--R diagram where this type dynamo operates are shown in \cite{renada2014} and in \cite{auriere2015} and are supported by the theoretical models in \cite{charbonnel2017}. They explain the so-called “first and second magnetic strips” on the giant branches with the convective envelope properties, namely the large convective turnover time. Together with the rotation in intermediate mass stars in these strips on the H--R diagram, these values of  $\tau_{conv}$ yield a Rossby number less than 1 and hence, conditions for an efficient $\alpha-\omega$ dynamo action. If a giant inside a magnetic strip engulfed a planet, it will gain additional rotation and will appear as an extremely magnetically active star. However, a gain of angular momentum could also come from the inside during the dredge-up phases at the bottom of the RGB and after the He-burning, i.e. early-AGB \citep{schroderrenada2022}.  Hence, M-giants with faster rotation and strong magnetic field that are outside the second magnetic strip could indicate for a planet engulfment episode.

\begin{figure}
    \centering
    \includegraphics[width=\columnwidth]{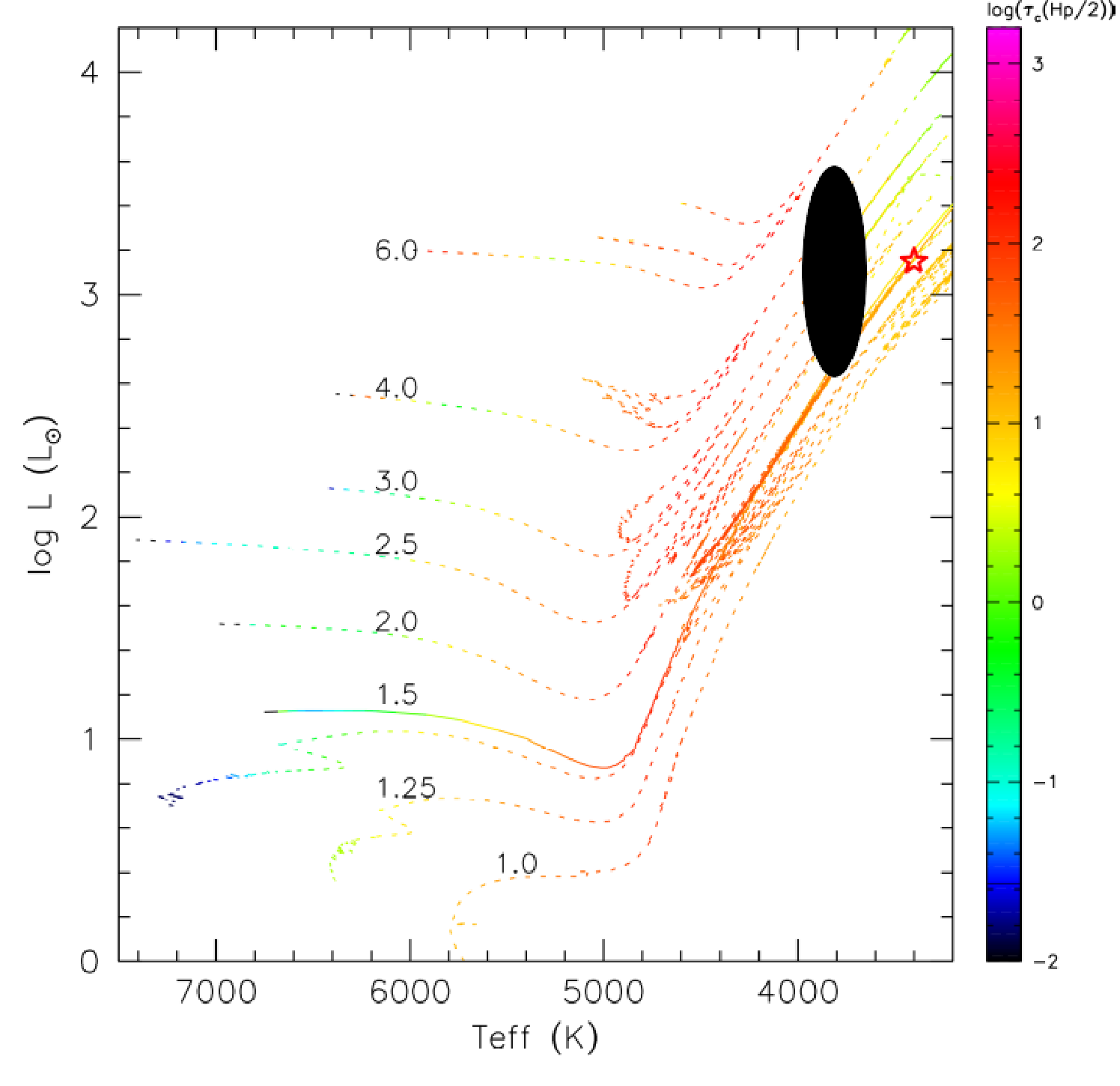}
    \caption{Situation of RZ~Ari on the H--R diagram. Colors stand for the convective turnover time at each point of the evolutionary tracks. Figure from \cite{renada2024}. The locus of the second "magnetic strip" is designated by an ellipse. }
    \label{fig:rzari-magnetic-strip}
\end{figure}

A typical example for magnetic activity triggered by planet engulfment is the M5 giant RZ Ari~\citep{renada2024}. It seems to be a single early AGB star outside the second magnetic strip (Figure~\ref{fig:rzari-magnetic-strip}). What is unusual in this star is its faster rotation ($V \sin i$ = $6\kms$, $P_{rot}$ = 530~days) and higher Lithium content (A(Li) = 1.2) that could not be explained by the theory of stellar evolution. A variable surface magnetic field on the order of a few Gauss up to 14 Gauss was Zeeman detected. It could not be explained in the term of $\alpha - \omega$ dynamo, because the  Rossby number is of the order of  20. However, a different type of dynamo, like $\alpha^2-\omega$ which is efficient for $R_o$ larger than 1 could operate there.  These results are also in a good agreement with the models for planet engulfment by \cite{privitera2016a}. Nevertheless, RZ~Ari is a semiregular pulsating star with a secondary period of about 480 days, and no relation between the pulsations and the magnetic field was found. The situation is similar in two other magnetic giants, EK~Boo and $\beta$~Peg \citep{georgiev2020}. RZ~Ari was also studied in search for the presence of large convective cells, because of the long periods in the magnetic field and photometric datasets of about 1200 days that could not be explained by rotational modulation \citep{renada2024}. No evidence for linear polarization was found in this star in 2015 when clear Zeeman detections and magnetic field of a few Gauss were observed. In the future, interferometry could shed more light on the surface structure and the existence of large convective cells. In any case, if large convective cells exist in RZ~Ari, an eventual local dynamo is not the dominant reason for the magnetic field and observed activity in this giant \citep{renada2024}.

Another giant candidate for planet engulfment is  IRAS 12556–7731 \citep{alcala2011}, where planet engulfment is also suspected to cause its relatively fast rotation and high lithium content. However, its magnetic field and activity are not yet studied.



\section*{3. The global picture. Conclusions. }
After more than a decade and a half of studies of the magnetic field in giants after the He-burning phase, two main mechanisms for the magnetic activity are evident: 1) dynamo of different types and 2) compression of a weak magnetic field as a result of shockwave propagation during pulsation. 
The dynamo should be initiated due to different reasons: evolutionary (the second magnetic strip) as it is in the models for the convective envelope and rotation, and $R_o$ by \cite{charbonnel2017}, or by the event of a planet engulfment that speeds up the surface rotation and could also initiate a dynamo. Since the engulfment could happen anywhere on the H--R diagram, the easiest way to find such stars is to look for magnetic giants outside the magnetic strips. Such a giant is already found -- RZ~Ari. It also has a high Lithium abundance in support of the planet engulfment hypothesis. While in the case of the second magnetic strip an $\alpha - \omega$ dynamo is possible, outside the strips (both the first and the second one) different kinds of dynamo could operate, depending on the properties of the convective envelope and the orbital momentum gained from the engulfed planet.

On the other hand, strong shocks during pulsations could compress the existing weak magnetic field. As a result, a magnetic field of a few Gauss is observed in some pulsating Mira-type stars \citep{lebre2014} and in post-AGB stars \citep{sabin2015, georgiev2023}. No magnetic field compression is observed in the studied early AGB stars (see Sect. 2.2). Perhaps the shockwaves are not sufficiently strong there. The absence of linear polarization in the spectra of $\beta$~Peg and RZ~Ari, two semi-regular pulsating variables, is also in support of this assumption \citep{georgiev2020, renada2024}.

It remains unclear, however, if a local dynamo like the one operating in Betelgeuse \citep{mathias2018} could operate in these intermediate mass AGB and post-AGB stars. Large convective cells are observed in the 2 M$_{\sun}$ AGB star $\pi^1$~Gru by means of interferometry \citep{paladini2018}. However, in our spectropolarimetric studies we do not detect linear polarization in the few AGB stars we observed. In the case of RZ~Ari, the star from our selection studied in the most detail, simultaneous linear and circular polarization observations in 2015 did not confirm a local dynamo operation. However, long periods of the order of 1200 days exist in the magnetic field and in the photometry datasets \citep{renada2024}. These periods could be explained by the lifetime of giant convective cells, according to \citep{freytag17}. If large convective cells exist in these giants, then we can assume that they are not as large as in the supergiant Betelgeuse, and so if a local dynamo does operate there, it is of smaller efficiency and does not dominate the magnetic field we detect. In the future, interferometric observations of these giants could reveal their surface structure. Further interferometric and spectropolarimetric studies are highly desired. 
 
Further studies could expand or modify this first knowledge on the mechanisms for MF and activity in ABG and post-AGB stars and their magnetic properties. However, such studies will take a decade or even more. Nevertheless, they deserve the efforts, because the magnetic activity in these stars is poorly studied and the results discussed above are just the beginning of an exiting new era in magnetism in fairly evolved stars.

\section*{Acknowledgements}
We thank the TBL team for service observations during all these years. Thanks also to the French PNPS program for the observations after 2014, to the OPTICON for the observations in semesters 2008B and 2011B and to the Bulgarian NSF grant DSAB 02/3 for the observations in 2010. Part of the study was carried out under Bulgarian-French collaboration (program RILA) and Bulgarian NSF project DN 18/2. SG and AL thank the University of Montpellier for its support through the I-SITE MUSE project \textit{MAGEVOL}.




\end{document}